\documentclass[12pt]{article}

\usepackage{amsmath}
\usepackage{amssymb}
\usepackage{graphicx}
\newcommand{\AddrAHEP}{
  {\it AHEP Group, Instituto de F\'{\i}sica Corpuscular --
    C.S.I.C./Universitat de Val{\`e}ncia \\
    Edificio de Institutos de Paterna, Apartado 22085,
  E--46071 Val{\`e}ncia, Spain}}

\newcommand{\AddrLNF}{
  {\normalsize \it INFN, 
    Laboratori Nazionali di Frascati, 
    C.P. 13, I00044 Frascati, Italy}}

\newcommand{\AddrUTFSM}{
  {\it  Centro de Estudios Subat\'omicos(CES),
Universidad T\'ecnica Federico Santa Mar\'\i a,
Casilla 110-V, Valpara\'\i so, Chile}}

\def\gsim{\raise0.3ex\hbox{$\;>$\kern-0.75em\raise-1.1ex\hbox{$\sim\;$}}}
\def\lsim{\raise0.3ex\hbox{$\;<$\kern-0.75em\raise-1.1ex\hbox{$\sim\;$}}}

\begin{document}
\begin{titlepage}

\begin{flushright}
hep-ph/yymmnnn \\
IFIC/07-68\\
\end{flushright}
\vspace*{3mm}

\begin{center}
\textbf{{\large Leptoquarks: Neutrino masses and accelerator phenomenology}}
\\[10mm]
D. Aristizabal Sierra$^{a,b}$, M. Hirsch$^a$ and 
S. G. Kovalenko$^c$
\vspace{0.3cm}\\
$^a$\AddrAHEP.\vspace{0.3cm}\\
$^b$\AddrLNF.\vspace{0.3cm}\\
$^c$\AddrUTFSM.\vspace{0.3cm}

\end{center}

\begin{abstract}
Leptoquark-Higgs interactions induce mixing between leptoquark states with
different chiralities once the electro-weak symmetry is broken. In such LQ
models Majorana neutrino masses are generated at 1-loop order. Here we
calculate the neutrino mass matrix and explore the constraints on the
parameter space enforced by the assumption that LQ-loops explain current
neutrino oscillation data. LQs will be produced at the LHC, if their masses
are at or below the TeV scale. Since the fermionic decays of LQs are
governed by the same Yukawa couplings, which are responsible for the
non-trivial neutrino mass matrix, several decay branching ratios of LQ
states can be predicted from measured neutrino data. Especially interesting
is that large lepton flavour violating rates in muon and tau final states
are expected. In addition, the model predicts that, if kinematically
possible, heavier LQs decay into lighter ones plus either a standard model
Higgs boson or a $Z^0/W^{\pm}$ gauge boson. Thus, experiments at the LHC
might be able to exclude the LQ mechanism as explanation of neutrino data.
\end{abstract}
\end{titlepage}

\section{Introduction}
\label{sec:int}

Leptoquarks (LQs) appear in many extensions of the standard model. First
discussed in the classic papers by Pati \& Salam \cite{Pati:1974yy} and
Georgi \& Glashow \cite{Georgi:1974sy}, LQs are a common ingredient to
grand unified theories \cite{Langacker:1980js}. \footnote{For a recent
example of a GUT model with light leptoquarks, see \cite{Dorsner:2005fq}.}
They can also appear in composite \cite{Schrempp:1984nj} as well as in
technicolour models \cite{Dimopoulos:1979es,Eichten:1979ah}. Also in
supersymmetric models with R-parity violation scalar quarks have
leptoquark-like interactions \cite{Hall:1983id}. From a low-energy
point of view, however, LQs are best described in a ``model-independent''
way, using a LQ Lagrangian based only on the minimal assumptions of (a)
renormalizability and (b) standard model (SM) gauge invariance
\cite{Buchmuller:1986zs}. An exhaustive list of limits on such LQs from
low-energy experiments can be found, for example, in \cite{Davidson:1993qk}.

Direct searches so far have not turned up any evidence for LQs
\cite{Chiarelli:2005dj}. The best limits on pair produced LQs currently
come from the D0 \cite{Abazov:2007bs} and CDF \cite{Acosta:2005ge}
experiments at the Tevatron. These typically give limits on LQ masses
in the ballpark of $m_{LQ} \gsim (200-250)$ GeV, depending mainly on the
final state decay branching ratios and on the lepton-quark generation,
to which the LQ state couples. Considerably more stringent limits are
expected from the LHC experiments. Depending on the accumulated luminosity,
the LHC should be able to find LQs up to masses of order of
$m_{LQ} \sim (1.2-1.5)$ TeV \cite{Kramer:2004df}.

Solar \cite{sno}, atmospheric \cite{Fukuda:1998mi} and reactor \cite{kamland}
neutrino oscillation experiments have firmly established that neutrinos
have mass and non-trivial mixing between different generations. In the
SM neutrinos are massless. However, non-zero neutrino masses can easily
be generated and the literature is abound in neutrino mass models
\cite{Valle:2006vb}. Certainly the most popular way to generate neutrino
masses is the seesaw mechanism
\cite{Minkowski:1977sc,seesaw,MohSen,Schechter:1980gr}, countless variants
exist. However, it is also conceivable that the scale of lepton number
violation is near - or at - the electro-weak scale. To mention a few
examples, there are supersymmetric models with violation of R-parity
\cite{Hall:1983id,hirsch:2000ef}, models with Higgs triplets
\cite{Schechter:1980gr} or a combination of both
\cite{AristizabalSierra:2003ix}. Also purely radiative models have been
discussed in the literature, both with neutrino masses at 1-loop
\cite{Zee:1980ai,AristizabalSierra:2006ri} or at 2-loop
\cite{Nieves:1981tv,Zee:1985id,Babu:2002uu,AristizabalSierra:2006gb} order.
Radiative mechanisms might be considered especially appealing, since
they generate small neutrino masses automatically, essentially due to
loop suppression factors.

In this paper, we study the generation of neutrino masses due to loops
involving light leptoquarks, in a model with non-zero leptoquark-Higgs
interactions \cite{Hirsch:1996qy}. LQ-Higgs interactions lead to mixing
between LQs of different chiralities (and lepton number) once electro-weak
symmetry is broken and thus can contribute non-trivially to the Majorana
neutrino mass matrix at 1-loop level. \footnote{This is not an entirely
new subject. Majorana neutrino masses due to loops involving coloured
scalars (and vectors) have been discussed first in \cite{Nieves:1981tv}.
However, because in \cite{Nieves:1981tv} it was assumed that these particles
have masses $M \ge 10^8$ GeV, their contribution to ${\cal M}_{\nu}$ was
deemed negligible.} As discussed below, the peculiar structure of
leptonic mixing, observed in neutrino oscillation experiments, enforces
a number of constraints on the LQ parameter space. The main result
of our current work is, that these constraints can be used to make
definite predictions for different decay branching ratios of several
LQ states. Therefore, the hypothesis that LQ loops are responsible for
the generation of neutrino mass is testable at the LHC, if LQs have
masses of the order of ${\cal O}(1)$ TeV.

Before proceeding, a few more comments on LQs might be in order. First,
for the LQ model to be able to explain neutrino data, non-zero LQ-Higgs
interactions are essential. Limits on these couplings, on the other hand,
can be derived from low-energy data such as, for example, pion decay
\cite{Hirsch:1996qy}. Especially stringent are limits from neutrinoless
double beta decay \cite{Hirsch:1996ye} and from the decay
$K^0 \rightarrow e^{\pm}\mu^{\mp}$ \cite{Kolb:1997rb}. However, as we
will discuss below, the small neutrino masses themselves are up to now
the most sensitive low-energy probe of LQ-Higgs mixing terms.

Second, it should be mentioned that LQ loops as a source of neutrino mass
have been discussed previously in \cite{Mahanta:1999xd}. We will improve
upon this work in several aspects: (i) We will present neutrino mass
formulas containing all possible LQ loops, while in \cite{Mahanta:1999xd}
only down-type quark loops were considered. (ii) \cite{Mahanta:1999xd}
concentrated on upper limits on LQ parameters, which can potentially be
derived from observed neutrino masses. We, on the other hand, identify
the regions of LQ parameters were the neutrino mass matrix is dominated
by LQ loops, thus providing a potential explanation of oscillation data.
And, lastly but most importantly, (iii) we discuss possible accelerator
tests of the LQ hypothesis of neutrino masses, to the best of our knowledge
for the first time in the literature.

Finally, it should also be mentioned that LQs can be either scalar or
vector particles. We consider only scalars in details. However, we note
that most of our results straightforwardly apply also for vector LQs.

The rest of this paper is organized as follows. In section (\ref{sec:lqdefs})
we define the leptoquark interactions, both with quark-lepton pairs and
with the SM Higgs boson and discuss the LQ mass matrices. In section
(\ref{sec:nuloops}) we calculate the 1-loop neutrino mass matrix in the
LQ model. Some particularly simple and interesting limits are defined and
discussed analytically. The typical ranges of LQ parameters, required to
explain current neutrino data, are explored. We then turn to the
phenomenology of LQs at future accelerators in section (\ref{sec:lqdecays}).
It is found that some fermionic LQ decays trace the measured neutrino
angles and thus can serve, in principle, as a test of the LQ model.
Next we discuss LQ decays to the SM Higgs and to gauge bosons. Higgs
(and $Z^0$) decays should occur, if kinematically possible, due to the
non-zero LQ mixing required to explain neutrino masses and thus form a
particularly interesting signal of the LQ model. We then close the paper
with a short summary.

\section{Leptoquark basics}
\label{sec:lqdefs}

\subsection{Scalar leptoquark Lagrangian}
\label{sec:lqlag}

The SM symmetries allow five scalar LQs. Table \ref{tab:LQcharges} shows
their $SU(3)_c\times SU(2)_L\times U(1)_Y$ quantum numbers, as well as
their standard baryon and lepton number assignments. LQs which couple
non-chirally are strongly constrained by low-energy data
\cite{Davidson:1993qk}. Thus, the states $S_0^L$ and $S_0^R$ (as well
as $S_{1/2}^L$ and $S_{1/2}^R$), which have the same SM quantum numbers,
but couple to (quark) doublets and singlets, respectively, are usually
assumed to be independent particles. Under these assumptions, the most
general Yukawa interactions (LQ-lepton-quark) induced by the new scalar
fields are given by \cite{Buchmuller:1986zs}
\begin{eqnarray}
  \label{eq:lqlqint}
  {\cal L}_{{\mbox\tiny{LQ}}-l-q} &=&
  \lambda_{S_0}^{(R)}\,\overline{u^c}
  P_R e \;S_{0}^{R\dagger}
  + \lambda_{\tilde{S}_{0}}^{(R)}
  \,\overline{d^c} P_R e \;\widetilde{S}_{0}^{R\dagger}
  + \lambda_{S_{1/2}}^{(R)}
  \,\overline{u} P_L l \;S_{1/2}^{R\dagger}
  \nonumber\\
  && + \lambda_{\tilde{S}_{1/2}}^{(R)}
  \,\overline{d} P_L l \;\tilde{S}_{1/2}^{\dagger}
  + \lambda_{S_{0}}^{(L)}
  \,\overline{q^c} P_L i\tau_2 l
  \;S_{0}^{L\dagger}
  + \lambda_{S_{1/2}}^{(L)}
  \,\overline{q} P_R i\tau_2 e \;S_{1/2}^{L\dagger}
  \nonumber\\
  && + \lambda_{S_1}^{(L)}
  \,\overline{q^c} P_L i\tau_2\;\widehat{S}_1^\dagger\; l
  + \mbox{h.c.}
\end{eqnarray}
Here we used the conventions of \cite{Davidson:1993qk}. Note that
eq.~(\ref{eq:lqlqint}) is written in one-generation notation. In
general, all $\lambda$'s are $3\times 3$ matrices in generation space.
$q$ and $l$ ($u$, $d$ and $e$) are the quark and lepton SM doublets
(singlets), $S^j_i$ are the scalar LQs with the weak isospin $i=0,1/2,1$
coupled to left-handed ($j=L$) or right-handed ($j=R$) quarks respectively.
Thus, in total eq.~(\ref{eq:lqlqint}) contains seven LQ fields. We have
also defined $\widehat{S}_1 = \tau \cdot S_1$.

\begin{table}[t]
  \centering
  \renewcommand{\arraystretch}{1.1}
  \renewcommand{\tabcolsep}{0.3cm}
  \begin{tabular}{|l|c|c|c|c|c|c|}
    \hline
    LQ & $SU(3)_c$ & $SU(2)_L$ & $Y$ &
    $Q_{\mbox{\tiny{em}}}$ & L & B
    \\\hline
    $S_0$ & $\mathbf{3}$ & $\mathbf{1}$ & -2/3 & -1/3 & 1 & 1/3\\
    $\widetilde{S}_0$ & $\mathbf{3}$ & $\mathbf{1}$
    & -8/3 & -4/3 & 1 & 1/3\\
    $S_{1/2}$ & $\mathbf{3^*}$ & $\mathbf{2}$ &
    -7/3 & (-2/3,-5/3) & 1 & -1/3 \\
    $\widetilde{S}_{1/2}$ & $\mathbf{3^*}$ &
    $\mathbf{2}$ & -1/3 & (1/3,-2/3) & 1 & -1/3 \\
    $S_{1}$ & $\mathbf{3}$ & $\mathbf{3}$ &
    -2/3 & (2/3,-1/3,-4/3) & 1 & 1/3 \\
    \hline
  \end{tabular}
  \caption[Standard model quantum numbers of the scalar
  leptoquarks]{Standard model quantum numbers of the
    scalar leptoquarks. The indices 0, 1/2, 1 indicate
    the weak isospin. The weak hypercharge is normalized 
    according to $Y=2(Q_{em}-T_3)$.}
  \label{tab:LQcharges}
\end{table}

The most general renormalizable and gauge invariant
scalar LQ interactions with the SM Higgs doublet ($H$)
are described by the scalar potential
\cite{Hirsch:1996qy}
\begin{eqnarray}
  \label{eq:LQhiggsint}
  V &=&
  h_{S_0}^{(i)}H i \tau_2 \;\widetilde{S}_{1/2}\;S_0^i
  +
  h_{S_1}H i \tau_2 \; \widehat{S}_{1}
  \;\widetilde{S}_{1/2}
  +
  Y_{S_{1/2}}^{(i)}
  \left(
    H i \tau_2 S^i_{1/2}
  \right)
  \left(
    \widetilde{S}^\dagger_{1/2} H
  \right)
  \nonumber\\
  && + Y_{S_1}
  \left(
    H i \tau_2\widehat{S}_1^\dagger H
  \right) \widetilde{S}_0
  %\nonumber\\
  + \kappa_S^{(i)}
  \left(
    H^\dagger \widehat{S}_1 H
  \right)S_0^{i\dagger}
  \nonumber\\
  && - (M_\Phi^2 - g_\Phi^{(i_1 i_2)}H^\dagger H)
  \Phi^{i_1\dagger}\Phi^{i_2} + \mbox{h.c.}
\end{eqnarray}
Here $\Phi^i$ is a cumulative notation for all scalar
LQ fields with $i=L,R$ (the same for $i_{1,2}$). The
diagonal mass terms $M_\Phi^2\Phi^\dagger\Phi$ can be
generated by spontaneous breaking of the fundamental
underlying symmetry down to the electroweak gauge
group at some high-energy scale.
The subsequent electroweak symmetry breaking produces
non-diagonal LQ mass terms which, in addition to the
diagonal terms given in eq.~(\ref{eq:LQhiggsint}), define
the LQ squared-mass matrices. These will be discussed next. 

%{SK_} I added this phrase which explains the role of these terms 
%for the Majorana neutrino mass generation
It is important to note, that the first two terms of the scalar 
potential in eq.~(\ref{eq:LQhiggsint}) violate total lepton number 
by two units $\Delta L =2 $ and, therefore, generate Majorana neutrino 
masses after electro-weak symmetry breaking. 
\footnote{Note that an extension of the scalar potential in 
eq.~(\ref{eq:LQhiggsint}) to include the LQ trilinear self-interaction 
terms has been discused in \cite{SchmidtKovalenko}. Such terms, 
however, introduce violation of baryon number and, hence, proton 
decay, for details see \cite{SchmidtKovalenko}. Therefore we will 
not consider LQ self-interaction terms here.}
In the limit where $h_{S_0}^{(R)}$ and $h_{S_1}$ vanish, neutrino 
masses vanish as well.

\subsection{Scalar leptoquark mass spectrum}
\label{sec:lqmassspectrum}
There are four squared-mass matrices which determine the masses of LQs
with the same electric charge ($Q=-1/3, -2/3, -4/3, -5/3$). In the
interaction eigenstate basis, defined by
$S_{-1/3}=(S_0^L, S_0^R, \widetilde{S}_{1/2}^\dagger, S_1)$,
$S_{-2/3}=(\widetilde{S}_{1/2},S^L_{1/2}, S^R_{1/2},S_1^\dagger)$,
$S_{-4/3}=(\widetilde{S}_0,S_1)$ and
$S_{-5/3}=(S_{1/2}^L,S_{1/2}^R)$,
the squared-mass matrices read
\begin{equation}
  \label{eq:q1/3lqmassm}
  {\cal  M}^2_{-1/3}=
  \begin{pmatrix}
    \overline{M}^2_{S_0^L} & g^{(LR)}_{S_0} v^2      &
    h_{S_0}^L v & \kappa_S^{(L)} v^2 \\
    \cdot & \overline{M}^2_{S_0^R} & h_{S_0}^R v     &
    \kappa_S^{(R)} v^2 \\
    \cdot & \cdot & \overline{M}^2_{\tilde{S}_{1/2}} &
    h_{S_1} v \\
    \cdot & \cdot & \cdot & \overline{M}^2_{S_1}
  \end{pmatrix} ,
\end{equation}
\begin{equation}
  \label{eq:q2/3lqmassm}
  {\cal  M}^2_{-2/3} =
  \begin{pmatrix}
    \overline{M}^2_{\tilde{S}_{1/2}} & Y^L_{S_{1/2}} v^2 &
    Y^R_{S_{1/2}} v^2 & \sqrt{2} h_{S_1} v \\
    \cdot & \overline{M}^2_{S^L_{1/2}}                   &
    g^{(LR)}_{S_{1/2}} v^2 & 0 \\
    \cdot & \cdot & \overline{M}^2_{S^R_{1/2}} & 0 \\
    \cdot & \cdot & \cdot & \overline{M}^2_{S_1}
  \end{pmatrix} ,
\end{equation}
\begin{equation}
  \label{eq:q4/3lqmassm}
  {\cal  M}^2_{-4/3} =
  \begin{pmatrix}
    \overline{M}^2_{\tilde{S}_0} & \sqrt{2} Y_{S_1} v^2\\
    \cdot                       &
    \overline{M}^2_{S_1}
  \end{pmatrix} ,
\end{equation}
and
\begin{equation}
  \label{eq:q5/3lqmassm}
  {\cal  M}^2_{-5/3} =
  \begin{pmatrix}
    \overline{M}^2_{S^L_{1/2}} & - g^{(LR)}_{S_{1/2}} v^2\\
    \cdot & \overline{M}^2_{S^R_{1/2}}
  \end{pmatrix} .
\end{equation}
Here $\overline{M}^2_\Phi= M_\Phi^2 - g_{\Phi}v^2$ and only the
elements above the diagonal have been written since the matrices are
symmetric. $v$ is the SM Higgs vacuum expectation value,
$v^2 = (2\sqrt{2}G_F)^{-1}$. The mass eigenstate basis is defined as
\begin{equation}
  \label{eq:massandinteractionbasis}
  (\widehat{S}_Q)_i = R^{Q}_{ij}\;(S_Q)_j ,
\end{equation}
where $R^Q$ is a rotation matrix. The diagonal squared-mass matrices are
found in the usual way:
\begin{equation}
  \label{eq:LQdiagmassmatrices}
  \left(
    {\cal  M}^2_Q
  \right)_{\mbox{\tiny{diag}}}
  =
  R^Q {\cal  M}^2_Q (R^Q)^T .
\end{equation}
Phenomenological implications of the LQ interactions given in eqs.
(\ref{eq:lqlqint}) and (\ref{eq:LQhiggsint}) have to be derived in
terms of the mass eigenstates. For the LQs with charge $Q=-4/3$ and
$Q=-5/3$, simple analytical expressions for the eigenvalues and
rotation angle can be found. These are given by
\begin{equation}
\label{eq:LQeigs_sim}
M^2_{1,2} = \frac{1}{2}\Big(M^2_{11}+M^2_{22}-
\sqrt{4 M^4_{12} + (M^2_{11}-M^2_{22})^2}\Big),
\end{equation}
and
\begin{equation}
\label{eq:LQang_sim}
\tan 2\theta_{12} = \frac{2 M^2_{12}}{M^2_{11}-M^2_{22}}.
\end{equation}
Here, $M^2_{11}$, $M^2_{22}$ and $M^2_{12}$ stand symbolically for the
corresponding entries in the mass matrices eqs (\ref{eq:q4/3lqmassm})
and (\ref{eq:q5/3lqmassm}). For LQs of charge $Q=-1/3,-2/3$ we will
diagonalize the mass matrices numerically below. However, the following
approximate expressions are useful for an analytical estimate of
parameters. The rotation matrices which relate the interaction and
mass eigenstates can be parametrized by six rotation angles, namely,
\begin{equation}
  \label{eq:rotlq1323}
  R^Q=R^Q(\theta_{34})R^Q(\theta_{24})R^Q(\theta_{14})
  R^Q(\theta_{23})R^Q(\theta_{13})R^Q(\theta_{12}).
\end{equation}
In the limit where the off-diagonal entries in the mass matrices
eqs (\ref{eq:q1/3lqmassm}) and (\ref{eq:q2/3lqmassm}) are smaller than
the difference between the corresponding diagonal ones, it is possible
to find approximate expressions for the rotation angles also in this
more complicated case. As discussed in the next section, for the
neutrino masses the most relevant angles are $\theta^{Q=2/3}_{34}$,
$\theta^{Q=1/3}_{34}$ and $\theta^{Q=1/3}_{13}$. For the angles in
the $Q=1/3$ case on can use eq. (\ref{eq:LQang_sim}) as an estimate,
with obvious replacements of indices. For the angle $\theta^{Q=2/3}_{34}$,
however, since the relevant $M^2_{34} =0$ in the mass basis, a more
complicated expression results:
\begin{equation}
\label{eq:rot34app}
\theta^{Q=2/3}_{34}\simeq -\frac{\sqrt{2}Y^R_{S_{1/2}}h_{S_1}v^3}
                {(\overline{M}^2_{S^R_{1/2}}-\overline{M}^2_{\tilde{S}_{1/2}})
                 (\overline{M}^2_{S_1} -\overline{M}^2_{S^R_{1/2}})}.
\end{equation}
Eq. (\ref{eq:rot34app}) is exact in the limit $Y^L_{S_{1/2}} =
g^{(LR)}_{S_{1/2}}=0$. It remains a reasonable (factor-of-two) estimate
as long as $Y^L_{S_{1/2}},g^{(LR)}_{S_{1/2}}\le Y^R_{S_{1/2}}h_{S_1}/v$,
and all $Q=2/3$ LQ mixing angles are small numbers.

\section{Neutrino masses from Leptoquark loops}
\label{sec:nuloops}

\subsection{Analytical formulas}

LQ-$q\nu$ Yukawa interactions can be derived directly from the
Lagrangian~(\ref{eq:lqlqint}). In the interaction eigenstate
basis, they have the following form
\begin{equation}
  \label{eq:unulqLag}
  {\cal L}_{LQ-u\nu}=\lambda_{S_{1/2}}^R\bar{u}P_L\nu
  (S_{1/2}^R)^\dagger_{-2/3}
  + \lambda_{S_1}^L\bar{u^c}P_L\nu(S_1)^\dagger_{-2/3}
  + \mbox{h.c.}
\end{equation}
and
\begin{equation}
  \label{eq:dnulqlag}
  {\cal L}_{LQ-d\nu}=
  \lambda_{\tilde{S}_{1/2}}^R\bar{d}P_L\nu
  (\tilde{S}_{1/2}^R)^{\dagger}_{-1/3}
  - \lambda_{S_0}^L\bar{d^c}P_L\nu(S_0^L)^\dagger_{1/3}
  + \lambda_{S_1}^{L}\bar{d^c}P_L\nu(S_1)^\dagger_{1/3}
  + \mbox{h.c.}
\end{equation}
\begin{figure}[t]
  \centering
  \includegraphics{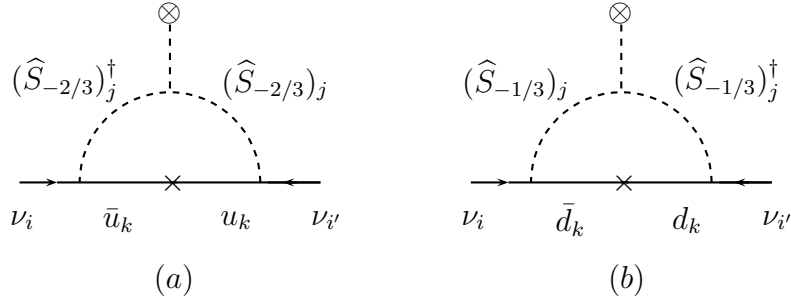}
  \caption[Feynman diagrams for Majorana neutrino masses]
  {Feynman diagrams for Majorana neutrino masses.
    Diagram $(a)$ [$(b)$] give contributions to the
    neutrino mass matrix from $u$-type [$d$-type]
    quarks loops.}
  \label{fig:LQloopsdiag}
\end{figure}
Rotating to the mass eigenstate basis, the non-trivial mixing among
LQs from different $SU(2)_L$ multiplets lead to neutrino Majorana masses
at one-loop order as shown in figure \ref{fig:LQloopsdiag}. A
straightforward calculation of the Majorana neutrino mass matrix
from these diagrams gives
\begin{equation}
  \label{eq:nmm1}
  {\cal M}_\nu={\cal M}^{\rm{up}}_\nu
  + {\cal M}^{\rm{down}}_\nu
\end{equation}
where the matrix ${\cal M}^{\rm{up}}_\nu$ from diagram $(a)$ reads
\begin{eqnarray}
  \label{eq:nmmup}
  \left(
    {\cal M}^{\rm{up}}_\nu
  \right)_{ii'}&=&
  \frac{3}{16\pi^2}
  \sum_{\substack{j=1\cdots 4\\k=u,c,t}} m_k
  B_0(0,m_k^2,M_j^2)R^{2/3}_{j3}R^{2/3}_{j4}\nonumber\\
  &&\times\left[
    (\lambda_{S_{1/2}}^R)_{ik}(\lambda_{S_1}^L)_{i'k}
    + (\lambda_{S_{1/2}}^R)_{i'k}(\lambda_{S_1}^L)_{ik}
  \right].
\end{eqnarray}
Here $R^{2/3}$ is the rotation matrix that diagonalizes the mass matrix
of $Q=-2/3$ LQs, eq.~(\ref{eq:q2/3lqmassm}), and $B_0(0,m_k^2,M_j^2)$
is a Passarino-Veltman function \cite{Passarino:1978jh}. The matrix
${\cal M}_\nu^{\rm{down}}$ from diagram $(b)$ can be written as
\begin{eqnarray}
  \label{eq:nmmdown}
  \left(
    {\cal M}^{\rm{down}}_\nu
  \right)_{ii'}&=&
  \frac{3}{16\pi^2}
  \sum_{\substack{j=1\cdots 4\\k=d,s,b}} m_k
  B_0(0,m_k^2,M_j^2)R^{1/3}_{j3}\nonumber\\
  &&\times
  \left\{
    R^{1/3}_{j4}
    \left[
      (\lambda_{\tilde{S}_{1/2}}^R)_{ik}
      (\lambda_{S_1}^L)_{i'k}
      + (\lambda_{\tilde{S}_{1/2}}^R)_{i'k}
      (\lambda_{S_1}^L)_{ik}
    \right]
%aqui
  \right.\nonumber\\
  &&+
  \left. R^{1/3}_{j1}
    \left[
      (\lambda_{\tilde{S}_{1/2}}^R)_{ik}
      (\lambda_{S_0}^L)_{i'k}
      + (\lambda_{\tilde{S}_{1/2}}^R)_{i'k}
      (\lambda_{S_0}^L)_{ik}
    \right]
  \right\} .
\end{eqnarray}
Here $R^{1/3}$ is the rotation matrix that diagonalizes the mass matrix
of the $Q=-1/3$ LQs given in eq.~(\ref{eq:q1/3lqmassm}). Note, that
in the limit of unmixed LQs, i.e. $R_{ij}=\delta_{ij}$, the neutrino
mass matrix vanishes.

The Passarino-Veltman function $B_0$ \cite{Passarino:1978jh} contains a
finite and an infinite part. However, since the LQ model does not have
a neutrino mass at tree-level there are no counter terms, which allow
to absorb infinities. The infinite parts of the $B_0$ functions therefore
must cancel among the different contributions in eq. (\ref{eq:nmmup}) and
eq. (\ref{eq:nmmdown}). Using the parameterization of the LQ rotation
matrices given in eq. (\ref{eq:rotlq1323}), we have checked that this
is indeed the case. The resulting formula can be expressed as a sum
of {\em differences} of $B_0$ functions only, thus cancelling all infinities.
Since the coefficients in these formulas are rather lengthy (and of
little use), we will not give them explicitly.

\noindent 
Diagonalizing eq. (\ref{eq:nmm1}) gives the neutrino 
masses and mixing angles,
\begin{equation}
\label{eq:defU}
U^T {\cal M}_\nu U = {\cal M}_\nu^{diag}.
\end{equation}
In standard parameterization $U$ is written as
\begin{equation}
\label{eq:paraU}
U= \left(\begin{array}{cccc}
         1 & 0 & 0 \cr
         0 & c_{23} & s_{23} \cr
         0 & -s_{23} & c_{23}
\end{array}\right)
\left(\begin{array}{cccc}
         c_{13} & 0 & s_{13}e^{-i\delta} \cr
         0 & 1 & 0 \cr
         -s_{13}e^{i\delta} & 0 & c_{13}
\end{array}\right)
\left(\begin{array}{cccc}
         c_{12} & s_{12} & 0 \cr
         -s_{12} & c_{12} & 0 \cr
          0 & 0 & 1
\end{array}\right) ,
\end{equation}
where $c_{ij}=\cos\theta_{ij}$ and $s_{ij}=\sin\theta_{ij}$ and $\delta$
is a CP-violating phase. Since we will consider only real parameters,
$\delta = 0, \pi$ and we have not written any Majorana phases in
eq. (\ref{eq:paraU}).

In general, the neutrino mass matrix receives contributions from diagrams
involving up-type ($u$-loops) and down-type ($d$-loops) quarks. In order
to find the eigensystem of eq. (\ref{eq:nmm1}) one has to solve a cubic
equation. However, much simpler analytical formulas can be derived, if
one particular loop dominates over all others. For example, in the limit
where only the top loop contributes to ${\cal M}_\nu$ one finds
$\mbox{Det}\;[{\cal M}_\nu]=0$, i.e. one of the three eigenvalues of the
mass matrix goes to zero. Note, that in this limit the model can produce
only a (normal) hierarchical neutrino spectrum.

Analytical expressions for the two non-zero neutrino masses can be
found easily in the limit $\mbox{Det}\;[{\cal M}_\nu]=0$. It is useful
to define two vectors in parameter space
\begin{eqnarray}
  \label{eq:vectorsRL}
  \mathbf{R} =
  \left[
    (\lambda^R_I)_1,(\lambda^R_I)_2,(\lambda^R_I)_3
  \right] ,
  \nonumber\\
  \mathbf{L} =
  \left[
    (\lambda^L_{I'})_1,(\lambda^L_{I'})_2,(\lambda^L_{I'})_3
  \right] .
\end{eqnarray}
Here, $(\lambda^{R,L}_{I,I'})_j = (\lambda^{R,L}_{I,I'})_{jk}$, with $j$
being the leptonic index, whereas we have suppressed for brevity the
hadronic index $k$. The indices $I$ and $I'$ stand symbolically for
$I=S_{1/2}$ and $I'=S_1$ if the top loop dominates, or $I=\widetilde{S}_{1/2}$
and $I'=S_1$ or $I'=S_0$, if one of the bottom loops dominates.
In terms of these vectors the two non-zero neutrino masses are given by
\begin{equation}
  \label{eq:neutrinomassesexactexp}
  m_{\nu_{2,3}}= {\cal F} (|\mathbf{R}\cdot \mathbf{L}| \mp
  |\mathbf{R}| |\mathbf{L}|) ,
\end{equation}
where ${\cal F}$ is given by
\begin{equation}
\label{eq:defF}
{\cal F} =   \frac{3}{16\pi^2}
  \sum_{\substack{j=1\cdots 4}} m_k
  B_0(0,m_k^2,M_j^2)R^{Q}_{j3}R^{Q}_{js},
\end{equation}
$Q=1/3,2/3$ and $s=1,4$, depending on which contribution to ${\cal M}_{\nu}$
is most important. The ratio between the solar and the atmospheric scale is
thus simply given by
\begin{equation}
  \label{eq:ratio}
  R\equiv \frac{\Delta m^2_{12}}{\Delta m^2_{23}}\simeq
  \left(
    \frac{|\mathbf{R}\cdot \mathbf{L}| - |\mathbf{R}| |\mathbf{L}|}
    {|\mathbf{R}\cdot \mathbf{L}| + |\mathbf{R}| |\mathbf{L}|}
  \right)^2 .
\end{equation}
Note that $R$ is independent of ${\cal F}$, i.e. independent of LQ masses
and mixings. Relations among neutrino mixing angles and the Yukawa
couplings can be found by using the eigenvalue equation for the massless
neutrino \cite{Babu:2002uu,AristizabalSierra:2006gb},
\begin{equation}
  \label{eq:eigeqtopcase}
  {\cal M}_\nu\,v_0=0
\end{equation}
where the eigenvector $v_0$ is given by
\begin{equation}
\label{eq:eigenvectortopcase}
v_0^T=\frac{(1,-\epsilon,\epsilon')}
{\sqrt{\epsilon^2 + \epsilon'^2 + 1}} .
\end{equation}
Solving eq.~(\ref{eq:eigeqtopcase}) yields the result
\begin{equation}
  \label{eq:epsandepsp}
  \epsilon=\frac{m_{12}m_{33}-m_{13}m_{23}}
  {m_{22}m_{33}-m_{23}^2},
  \quad
  \epsilon'=\frac{m_{12}m_{23}-m_{13}m_{22}}
  {m_{22}m_{33}-m_{23}^2},
\end{equation}
where $m_{ij}$ are the entries of the neutrino mass matrix $\cal{M}_{\nu}$. 
Interestingly, eq. (\ref{eq:epsandepsp}) can be expressed in terms of
neutrino angles only. For a normal hierarchical spectrum, i.e.
$m_{\nu_{1,2,3}}\simeq(0,m,M)$, where $M$ ($m$) stands for the
atmospheric (solar) mass scale one obtains
\begin{eqnarray}
  \label{eq:epsepspmixa1}
  \epsilon&=&\tan\theta_{12}
  \frac{\cos\theta_{23}}{\cos\theta_{13}}
  +\tan\theta_{13}\sin\theta_{23},
  \\
  \label{eq:epsepspmixa2}
 \epsilon'&=&\tan\theta_{12}
 \frac{\sin\theta_{23}}{\cos\theta_{13}}
 -\tan\theta_{13}\cos\theta_{23}.
\end{eqnarray}
On the other hand, the expressions for $\epsilon$ and $\epsilon'$
eq. (\ref{eq:epsandepsp}) depend on the entries in the neutrino mass
matrix which are determined by the LQ Yukawa couplings,
\begin{eqnarray}
  \label{eq:epsilonparam}
  \epsilon=&
  \frac{(\lambda_I^R)_3(\lambda_{I'}^L)_1
    -
    (\lambda_I^R)_1(\lambda_{I'}^L)_3}
  {(\lambda_I^R)_3(\lambda_{I'}^L)_2
    - (\lambda_I^R)_2(\lambda_{I'}^L)_3}\\
  \label{eq:epsilonparam1}
  \epsilon'=&\frac{(\lambda_I^R)_2
    (\lambda_{I'}^L)_1
    - (\lambda_I^R)_1(\lambda_{I'}^L)_2}
  {(\lambda_I^R)_3(\lambda_{I'}^L)_2
    - (\lambda_I^R)_2(\lambda_{I'}^L)_3}.
\end{eqnarray}
The above equations allow to relate the Yukawa couplings directly to
the measured neutrino angles. Note also, that current neutrino data
require both, $\epsilon$ and $\epsilon'$, to be non-zero.

\subsection{Neutrino data and parameter estimates}
\label{sec:neutrparamLQ}

Before discussing the constraints on LQ parameter space imposed by
neutrino physics, let us briefly recall that from neutrino oscillation
experiments two neutrino mass squared differences and two neutrino
angles are by now known quite precisely \cite{Maltoni:2004ei}. These are
the atmospheric neutrino mass, $\Delta m^2_{\rm Atm} = (2.0-3.2)$ [$10^{-3}$
eV$^2$], and angle, $\sin^2\theta_{\rm Atm} = (0.34-0.68)$, as well as the
solar neutrino mass $\Delta m^2_{\odot} = (7.1-8.9)$ [$10^{-5}$ eV$^2$],
and angle, $\sin^2\theta_{\odot} = (0.24-0.40)$, all numbers at 3 $\sigma$
c.l. For the remaining neutrino angle, the so-called Chooz
\cite{Apollonio:2002gd} or reactor neutrino angle $\theta_R$, a global fit
to all neutrino data \cite{Maltoni:2004ei} currently gives a limit of
$\sin^2\theta_R \le 0.04$ $@$ 3 $\sigma$ c.l.

Neutrino oscillation experiments have no sensitivity on the absolute
scale of neutrino masses, but the atmospheric data requires that at
least one neutrino has a mass larger than $M\equiv m_{\nu}^{\rm Atm}\gsim 50$
meV. The minimal size of LQ Yukawa couplings and LQ-mixing, required to
explain such a neutrino mass, can be estimated from
eq. (\ref{eq:neutrinomassesexactexp}). Parameterizing the rotation
matrices as in eq.(\ref{eq:rotlq1323}) we can estimate ${\cal F}$
as
\begin{equation}
\label{eq:F2}
{\cal F} \simeq  \frac{3}{16\pi^2} m_k \sin(2\theta_{3s})\Delta B_{3s}.
\end{equation}
Here, $\sin(2\theta_{3s})$ stands symbolically for the largest LQ
mixing angle, and 
$\Delta B_{ij}=B_0(0,m_k^2,m_i^2)-B_0(0,m_k^2,m_j^2)$. 
The finite part $B_0^f$ of $B_0$ is given by 
\begin{equation}
\label{eq:defDelB}
B_{0}^{f}(0,m_k^2,m_j^2)=\frac{m_k^2\log(m_k^2)-m_j^2\log(m_j^2)}{m_k^2-m_j^2}.
\end{equation}
The maximum allowed value of $|\Delta B_{ij}|$ for $m_{LQ}\le 1.5$ TeV is
$|\Delta B_{ij}|\simeq 3$ ($3.5$) for $m_k=m_{t}$ ($m_k=0$). With the
current central values for the quark masses \cite{Yao:2006px} we then
find the maximal value(s) of ${\cal F}$ as
\begin{eqnarray}
\label{eq:numF}\nonumber
{\cal F}^{\rm max} & \simeq & [ 4.9\hskip1mm : \hskip1mm 0.14 \hskip1mm
: \hskip1mm 4.1\cdot 10^{-2} \hskip1mm : \hskip1mm 3\cdot 10^{-3} \hskip1mm
: \hskip1mm 1.7\cdot 10^{-4} \hskip1mm : \hskip1mm 8\cdot 10^{-5} ]
\hskip2mm {\rm GeV}, \\
&{\rm for} & \hskip3mm t \hskip3mm : \hskip4mm b \hskip4mm : \hskip7mm c
\hskip9mm : \hskip7mm s \hskip6mm : \hskip7mm d \hskip9mm : \hskip7mm u
\end{eqnarray}
for maximal LQ mixing, i.e. $\sin(2\theta_{3s})=1/2$. For this value of
${\cal F}$ the minimum values for the Yukawa couplings required to explain
the atmospheric mass scale are then very roughly given by
\begin{eqnarray}
  \label{eq:lambdaUlb}
  (\lambda^R_{S_{1/2}})_{it}(\lambda^L_{S_{1}})_{i't}
   \gsim  5.1 \times 10^{-12} & \hskip5mm ,\hskip5mm & \hskip-3mm
  (\lambda^R_{{\tilde S}_{1/2}})_{ib}(\lambda^L_{S_{0,1}})_{i'b}
   \gsim  1.8 \times 10^{-10} \\
  (\lambda^R_{S_{1/2}})_{ic}(\lambda^L_{S_{1}})_{i'c}
   \gsim  6.0 \times 10^{-10} & \hskip5mm ,\hskip5mm &
  (\lambda^R_{{\tilde S}_{1/2}})_{is}(\lambda^L_{S_{0,1}})_{i's}
   \gsim  8.0 \times 10^{-9}\nonumber \\
  (\lambda^R_{{\tilde S}_{1/2}})_{id}(\lambda^L_{S_{0,1}})_{i'd}
    \gsim  1.5 \times 10^{-7} & \hskip5mm ,\hskip5mm &
  (\lambda^R_{S_{1/2}})_{iu}(\lambda^L_{S_{1}})_{i'u}
   \gsim  3.0 \times 10^{-7}\nonumber
\end{eqnarray}
Obviously, unless the $\lambda_{ik}$ follow a hierarchy inversely
proportional to the SM quark masses, third generation quark loops
give the by far largest contribution to the neutrino mass matrix.

We should compare the minimal values of eq. (\ref{eq:lambdaUlb}) with
the constraints coming from low-energy phenomenology. The most stringent
upper bounds for the first generation Yukawa couplings
($(\lambda_{S_{1/2}}^R)_{i1}$ and $(\lambda_{S_1}^L)_{i1}$) are currently
found from the upper limit on the lepton flavour violating process
$\mu\,\mbox{Ti}\to e\,\mbox{Ti}$ \cite{Davidson:1993qk,Gabrielli:2000te}
\begin{eqnarray}
  \label{eq:fgYub}
  (\lambda_{S_{1/2}}^R)_{11}(\lambda_{S_{1/2}}^R)_{21}&<&
  2.6\times 10^{-7}
  \left(
    \frac{M_j}{100\rm{GeV}}
  \right)^2,\nonumber\\
  (\lambda_{S_1}^L)_{11}(\lambda_{S_1}^L)_{21}&<&
  1.7\times 10^{-7}
  \left(
    \frac{M_j}{100\rm{GeV}}
  \right)^2.
\end{eqnarray}
Here $j$ labels the corresponding mass of the LQ eigenstate. Upper bounds
for the second (and third) quark generation couplings come from the charged
lepton flavour violating decay $\mu\to e\gamma$ and are given by
\begin{eqnarray}
  \label{eq:sgYub}
  (\lambda_{S_{1/2}}^R)_{12(13)}(\lambda_{S_{1/2}}^R)_{22(23)}&<&
  1.8\times 10^{-5}
  \left(
    \frac{M_j}{100\rm{GeV}}
  \right)^2,\nonumber\\
  (\lambda_{S_1}^L)_{12(13)}(\lambda_{S_1}^L)_{22(23)}&<&
  1.8\times 10^{-5}
  \left(
    \frac{M_j}{100\rm{GeV}}
  \right)^2.
\end{eqnarray}
Here, we have updated \cite{Davidson:1993qk} with the current experimental
upper limit on Br($\mu\to e\gamma$) \cite{Yao:2006px}.

Although eq.(\ref{eq:fgYub}) constrains a different combination of
left- and right-LQ couplings than eq.(\ref{eq:lambdaUlb}) we conclude
that, barring cases where some fine-tuned cancellation between different
LQ contributions occur, we expect that first generation quark loops can
not explain current neutrino data. Second and third generation LQ loops,
on the other hand, could both produce the observed neutrino masses,
consistent with all phenomenological constraints. However, considering the
hierarchy in $m_c/m_t \sim 8\cdot 10^{-3}$ and $m_s/m_b \sim 0.02$, from
now on we will concentrate on third (quark) generation LQs.
Note that, comparing eq.(\ref{eq:lambdaUlb}) with eq.(\ref{eq:sgYub})
one finds that the atmospheric mass scale can be generated consistent
with low-energy constraints for LQ mixing as small as $10^{-6}$ ($10^{-5}$)
in case of top-loops (bottom-loops). These numbers are significantly
smaller than constraints derived from other low-energy processes
\cite{Hirsch:1996ye,Kolb:1997rb}.

The observed large mixing angles in the neutrino sector require certain
ratios of Yukawa couplings to be non-zero. This can be most easily
understood as follows. One can use eqs (\ref{eq:defU}) and (\ref{eq:paraU})
to invert the problem and calculate the neutrino mass matrix in the
``flavour basis'' (in the basis where the charged lepton mass matrix is
diagonal). The resulting ${\cal M}_\nu$ in the general case is a
complicated function of the eigenvalues and mixing angles. However,
as first observed in \cite{Harrison:2002er}, the so-called tri-bimaximal
mixing pattern
\begin{equation}
\label{eq:UHPS}
U^{\rm HPS} =
\left(\begin{array}{cccc}
\sqrt{\frac{2}{3}} & \sqrt{\frac{1}{3}} & 0 \cr
- \frac{1}{\sqrt{6}} &  \frac{1}{\sqrt{3}} & - \frac{1}{\sqrt{2}} \cr
- \frac{1}{\sqrt{6}} &  \frac{1}{\sqrt{3}} & \frac{1}{\sqrt{2}}
\end{array}\right).
\end{equation}
is a good first-order approximation to the observed neutrino angles. In
case of hierarchical neutrinos ${\cal M}_\nu^{diag}=(0,m,M)$ it leads to
\begin{equation}
\label{eq:MHPS}
{\cal M}_\nu^{\rm HPS} =
\frac{1}{2}\left(\begin{array}{cccc}
0 &  0 &  0 \cr
0 &  M & -M \cr
0 & -M &  M
\end{array}\right) +
\frac{1}{3}\left(\begin{array}{cccc}
m &  m &  m \cr
m &  m &  m \cr
m &  m &  m
\end{array}\right).
\end{equation}
Comparing eq. (\ref{eq:MHPS}) with the index structure of eq.
(\ref{eq:nmmup}) and (\ref{eq:nmmdown}), one expects that
\begin{eqnarray}
\label{eq:estlamR}
(\lambda^{L}_I)_1(\lambda^{R}_I)_3+(\lambda^{L}_I)_3(\lambda^{R}_I)_1
&\ll &
(\lambda^{L}_I)_2(\lambda^{R}_I)_3+(\lambda^{L}_I)_3(\lambda^{R}_I)_2 \\
\label{eq:estlamA}
(\lambda^{L}_I)_2(\lambda^{R}_I)_3+(\lambda^{L}_I)_3(\lambda^{R}_I)_2
& \gsim &
(\lambda^{L}_I)_2(\lambda^{R}_I)_2-(\lambda^{L}_I)_3(\lambda^{R}_I)_3
\end{eqnarray}
for the couplings which give the largest contribution to ${\cal M}_\nu$.
Eq. (\ref{eq:estlamR}) is essentially due to smallness of the
reactor angle, while eq. (\ref{eq:estlamA}) follows from the observed
near-maximality of the atmospheric angle. Note that, if more than
one loop contributes to ${\cal M}_\nu$ of eq. (\ref{eq:nmm1}), $m_{\nu_1}
\ne 0$, but the ``large'' off-diagonal entry in the (2,3) element of
${\cal M}_\nu$ always requires $(\lambda^{L/R}_I)_2 \sim (\lambda^{L/R}_I)_3$,
for at least one LQ state. Finally, it should also be mentioned that
the smallness of solar versus atmospheric splitting requires that the
vectors $\mathbf{R}$ and $\mathbf{L}$, defined in eq. (\ref{eq:vectorsRL}),
are nearly aligned for {\em all} vectors contributing to ${\cal M}_{\nu}$,
compare eq. (\ref{eq:ratio}).

There are three different contributions to the neutrino mass matrix,
see eqs (\ref{eq:nmmup}) and (\ref{eq:nmmdown}). The top loop is
proportional to $\theta^{Q=2/3}_{34} \sim Y^R_{S_{1/2}}\cdot h_{S_1}$,
while the bottom loop is either proportional to $\theta^{Q=1/3}_{34}
\sim h_{S_1}$ or to $\theta^{Q=1/3}_{13} \sim h^L_{S_0}$. Lacking
a theoretical ansatz for these parameters, it is not possible
to predict which of these give the dominant contribution to the
neutrino mass matrix. However, since $m_b/m_t \sim 2 \%$,
the top-loop will be most important, if all LQ mixing angles (and
Yukawa couplings) are of similar size. We will refer to this case,
${\cal M}_\nu = {\cal M}_\nu^t$, as scenario I. $\theta^{Q=2/3}_{34}$,
on the other hand, can be much smaller than the corresponding angles
in the down-type loops in those parts of parameter space where
all relevant off-diagonal entries in the LQ mass matrices are small.
In this case ${\cal M}_\nu \simeq {\cal M}_\nu^b$ and we will refer to
this situation as scenario II (II.a: if $h^L_{S_0} \ll h_{S_1}/v$ and
II.b: if $h_{S_1}/v \ll h^L_{S_0}$).

\section{Leptoquark collider phenomenology}
\label{sec:lqdecays}

LQs, once produced, will decay almost instantanously. There are two
different sets of possible final states. In the current model, apart
from the usual decays into a quark and a lepton there are also vector
($W^{\pm}$ and $Z^0$) and scalar ($h^0$) final states, if kinematically
allowed. We will discuss first the fermionic decays.

\subsection{Fermionic LQ decays}
Fermionic decays of the LQ mass eigenstates are dictated by the Yukawa
interactions given in the Lagrangian~(\ref{eq:lqlqint}). Possible final
states can be either $\bar{\ell}_i u_k$, $\bar{\ell}_i d_k$,
$\bar{\nu}_i u_k$ or $\bar{\nu}_i d_k$. Most interesting, from the
phenomenological point of view, are final states with charged leptons,
since these allow to tag the flavour. Partial widths for two-body final
states can be calculated in a straightforward manner. For charged
lepton final states these are given by,
\begin{equation}
  \label{eq:br13}
  \Gamma[(\widehat{S}_{-1/3})_j\to\ell_i\bar{t^c}]=n^j_{-1/3}
  \left\{
    [(\lambda_{S_1}^L)_{i3} R^{1/3}_{j4}]^2
    + [(\lambda_{S_0}^L)_{i3} R^{1/3}_{j1}]^2
    + [(\lambda_{S_0}^R)_{i3} R^{1/3}_{j2}]^2
  \right\};
\end{equation}
\begin{equation}
  \label{eq:br23}
  \Gamma[(\widehat{S}_{-2/3})_j\to \ell_i \bar{b}]=n^j_{-2/3}
  \left\{
    [(\lambda_{\tilde{S}_{1/2}}^R)_{i3}R^{2/3}_{j1}]^2
    + [(\lambda_{S_{1/2}}^L)_{i3}R^{2/3}_{j2}]^2
  \right\};
\end{equation}
\begin{equation}
  \label{eq:br43}
  \Gamma[(\widehat{S}_{-4/3})_j\to \ell_i\bar{b^c}]=n^j_{-4/3}
  \left\{
    [(\lambda_{\tilde{S}_0}^R)_{i3}R^{4/3}_{j1}]^2
    + [(\lambda_{S_1}^L)_{i3}R^{4/3}_{j2}]^2
  \right\};
\end{equation}
\begin{equation}
  \label{eq:br53}
  \Gamma[(\widehat{S}_{-5/3})_j\to \ell_i\bar{t}]=n^j_{-5/3}
  \left\{
    [(\lambda_{S_{1/2}}^L)_{i3}R^{5/3}_{j1}]^2
    + [(\lambda_{S_{1/2}}^R)_{i3}R^{5/3}_{j2}]^2
  \right\}.
\end{equation}
Here, $n^j_{Q}$ is an overall constant given by
\begin{equation}
  \label{eq:defN}
 n^j_Q = \frac{3}{16\pi m_{S_j}}\Big[1-\frac{m_q^2+m_l^2}{m_{S_j}^2}\Big]
\lambda^{1/2}(m_{S_j}^2,m_q^2,m_l^2)
\end{equation}
with $\lambda^{1/2}(a,b,c)$ the usual phase space factor,
$\lambda(a,b,c) = (a+b-c)^2 - 4 ab$, and $m_{S_j}$, $m_q$ and $m_l$ the
corresponding LQ, quark and lepton masses.
Absolute values for the LQ widths can not be predicted. However,
minimal [maximal] values can be estimated from the atmospheric neutrino
mass scale [low-energy bounds]. Putting all parameters to their extreme
values fermionic widths could be as small [large] as ${\cal O}$(eV)
[${\cal O}$(MeV)].

In the above equation, we have written only the partial widths to top
and bottom quarks. Formulas for the lighter generations can be found
with straightforward replacements of indices. However, since the
widths, eqs (\ref{eq:br13}) - (\ref{eq:br53}), are not suppressed by
quark masses, our assumption that 3rd generation quark loops give the
dominant contribution to ${\cal M}_{\nu}$, can in principle be checked
experimentally. For example if
$Br(\widehat{S}_{-5/3} \rightarrow t + \sum_i l_i)/
Br(\widehat{S}_{-5/3} \rightarrow j + \sum_i l_i)$ and
$Br(\widehat{S}_{-4/3} \rightarrow b + \sum_i l_i)/
Br(\widehat{S}_{-4/3} \rightarrow j + \sum_i l_i)$, where $j$ stands
for any non-b jet, are larger than $m_c/m_t$, charm (and up) loops are
guaranteed to be sub-dominant. Similar tests can be devised for the
case of bottom loops.

Note that if the mixing between different LQs is small, as is generally
expected, the decays of some of the LQ states $\widehat{S}_{Q}$ are
controlled by {\em the same Yukawa couplings that determine the
non-trivial structure of the neutrino mass matrix}. This observation
forms the basis of the different decay pattern predictions discussed
below. However, one complication arises from the fact that we can not
predict if the decays of the lightest or one of the heavier of the LQ
states is dictated by the Yukawas fixed by neutrino physics. Again, in
the limit of small LQ mixing, this question can be decided experimentally,
in principle. Consider, for example, scenario I, ${\cal M}_{\nu} \simeq
{\cal M}_{\nu}^t$. The decays controlled by neutrino physics are those
governed by $\lambda_{S_{1/2}}^R$ and $\lambda_{S_{1}}^L$. These states
couple mainly to lepton doublets. Their components have the same diagonal
entries in the LQ mass matrices, we expect them to have similar masses.
These states should have very roughly
$\Gamma[(\widehat{S}_{-5/3})\to \sum \ell_i\bar{t}] \sim
\Gamma[(\widehat{S}_{-2/3})\to \sum \nu_i\bar{t}]$ and
$\Gamma[(\widehat{S}_{-4/3})\to \sum \ell_i b] \sim
\Gamma[(\widehat{S}_{-1/3})\to \sum \nu_i b]$. The other $Q=4/3,5/3$ mass
eigenstates mainly couple to singlet leptons, i.e. these state do not
decay to neutrinos. In what follows below, we will always assume that
the small mixing limit is realized and the LQ states relevant for the
experimental cross-checks can be identified.

Combining eqs (\ref{eq:ratio}) and (\ref{eq:estlamA}) with the decay
rates eqs (\ref{eq:br13}) - (\ref{eq:br53}), one can derive some
qualitative expectations for some ratios of branching ratios of
fermionic LQ decays. In general, the constraint from the large
atmospheric angle, plus the smallness of $R=\Delta m^2_{\odot}/
\Delta m^2_{\rm Atm}$ can only be full-filled if there are (at least)
two LQ states which have similar branching ratios to muonic and
tau final states. At the same time, these LQ states should have
less final states with electrons, essentially due to eq.
(\ref{eq:estlamR}) and the upper limit on the reactor angle.

Much more detailed predictions for fermionic decays of LQs can be made in
the explicit scenarios defined in the last section. We will first discuss
in some detail the results for scenario I,
${\cal M}_{\nu} = {\cal M}_{\nu}^t$.
For all figures presented in the following we have scanned the Yukawa
parameter space randomly, in such a way that all low-energy bounds are
obeyed. We then numerically diagonalized the resulting neutrino mass
matrices and checked for consistency with current neutrino oscillation
data. Different correlations among ratios of branching ratios with
the different pieces of neutrino data are then found.

\begin{figure}[t!]
  \centering
  \includegraphics[width=9cm,height=6cm]{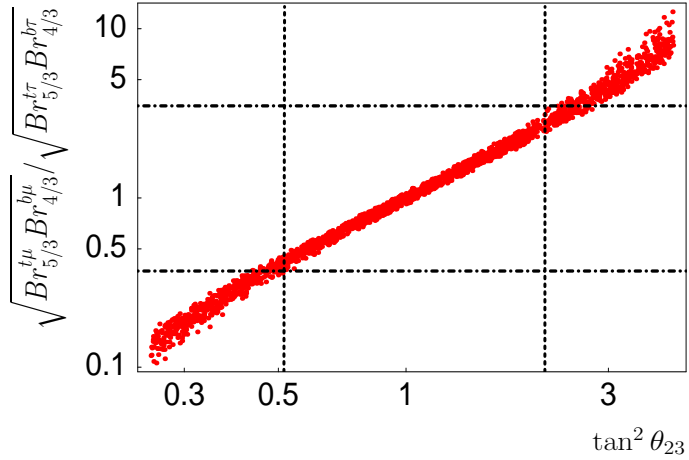}
  \caption[$\sqrt{Br_{5/3}^{t\mu}Br_{4/3}^{b\mu}}/
  \sqrt{Br_{5/3}^{t\tau}Br_{4/3}^{b\tau}}$
  versus $\tan^2\theta_{23}$.]
  {Ratio of decay branching ratios
    $\sqrt{Br_{5/3}^{t\mu}Br_{4/3}^{b\mu}}/
    \sqrt{Br_{5/3}^{t\tau}Br_{4/3}^{b\tau}}$ versus $\tan^2\theta_{23}$.
    Vertical lines indicate current $3\sigma$ range for
    $\tan^2\theta_{23}$ while horizontal lines determine the predicted
    range for this observable.}
  \label{fig:br_t23sq}
\end{figure}
Figure \ref{fig:br_t23sq} demonstrates that
$\sqrt{Br_{5/3}^{t\mu}Br_{4/3}^{b\mu}}/\sqrt{Br_{5/3}^{t\tau}Br_{4/3}^{b\tau}}$
is correlated with the atmospheric mixing angle. For the best fit point
value $\tan^2\theta_{23}=1$ one expects
$\sqrt{Br_{5/3}^{t\mu}Br_{4/3}^{b\mu}}\simeq
\sqrt{Br_{5/3}^{t\tau}Br_{4/3}^{b\tau}}$.
Using the current $3\sigma$ range for the atmospheric mixing angle this
observable can be predicted to lie within the interval [0.4,4.7].

\begin{figure}[t]
  \centering
  \includegraphics[width=9cm,height=6cm]{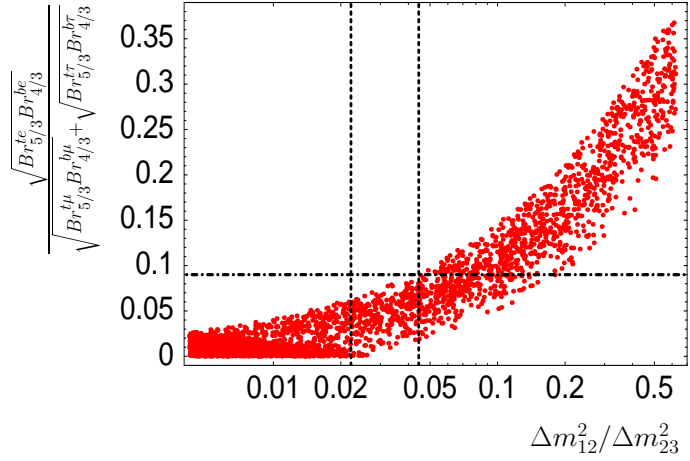}
  \caption[$\sqrt{Br_{5/3}^{t e}Br_{4/3}^{b e}}/
  (\sqrt{Br_{5/3}^{t\mu}Br_{4/3}^{b\mu}} +
  \sqrt{Br_{5/3}^{t\tau}Br_{4/3}^{b\tau}})$ versus $R$.]
  {Ratio of branching ratios
    $\sqrt{Br_{5/3}^{e\mu}Br_{4/3}^{e\mu}}/
    (\sqrt{Br_{5/3}^{t\mu}Br_{4/3}^{b\mu}} +
    \sqrt{Br_{5/3}^{t\tau}Br_{4/3}^{b\tau}})$ versus $R$.
    Vertical lines indicate current $3\sigma$ limits on $R$
    whereas the horizontal line shows the upper bound for this
    observable.}
  \label{fig:brcom_rat}
\end{figure}
\begin{figure}[h!]
  \centering
  \includegraphics[width=9cm,height=7cm]{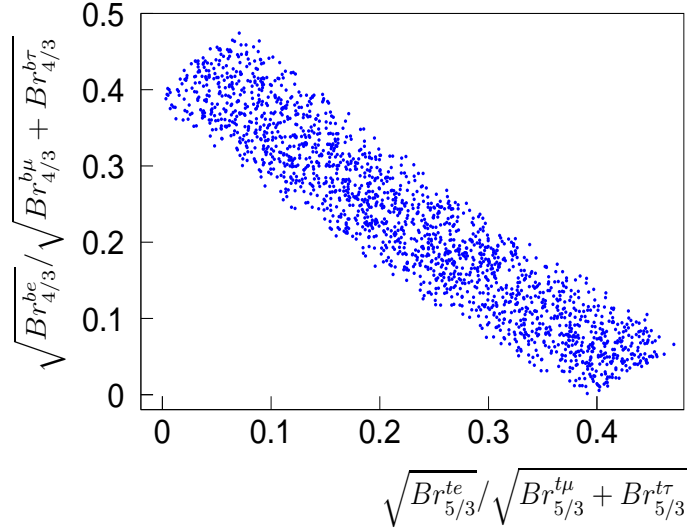}
  \caption[$\sqrt{Br_{4/3}^{be}}/\sqrt{Br_{4/3}^{b\mu}+Br_{4/3}^{b\tau}}$
  versus $\sqrt{Br_{5/3}^{te}}/\sqrt{Br_{5/3}^{t\mu}+Br_{5/3}^{t \tau}}$.]
  {Ratio of decay branching ratios
    $\sqrt{Br_{4/3}^{be}}/\sqrt{Br_{4/3}^{b\mu}+Br_{4/3}^{b\tau}}$
    versus $\sqrt{Br_{5/3}^{te}}/\sqrt{Br_{5/3}^{t\mu}+Br_{5/3}^{t \tau}}$.}
  \label{fig:b43com_b53com}
\end{figure}
We have found that there exists an upper bound on the ratio of
branching ratios
\begin{equation}
  \label{eq:b4353comrat}
  \frac{\sqrt{Br_{5/3}^{t e}Br_{4/3}^{b e}}}
  {\sqrt{Br_{5/3}^{t\mu}Br_{4/3}^{b\mu}} +
    \sqrt{Br_{5/3}^{t\tau}Br_{4/3}^{b\tau}}}
  \lesssim 9\times 10^{-2}
\end{equation}
which can be derived from the ratio $R=\Delta m^2_{\odot}/\Delta m^2_{\rm Atm}$
as shown in figure \ref{fig:brcom_rat}. This bound shows that the product
of branching ratios $Br_{5/3}^{te} Br_{4/3}^{be}$ is expected to be nearly
two orders of magnitude smaller than the sum of
$Br_{5/3}^{t \mu} Br_{4/3}^{b \mu}$ and $Br_{5/3}^{t\tau} Br_{4/3}^{b\tau}$.

Individual values for electron final state decay branching ratios are
shown in figure~\ref{fig:b43com_b53com}. It can be seen that the smallness
of $Br_{5/3}^{te} Br_{4/3}^{be}$ can be due to the smallness of either
$Br_{4/3}^{be}$ or $Br_{5/3}^{te}$. This implies that for one of the two
LQ eigenstates ($Q=4/3$ and $Q=5/3$) electron final states could be
as large as $\sim 20 \%$, but only if the other LQ state shows a very much
supressed branching ratio to electrons.

\begin{figure}[t!]
  \centering
  \includegraphics{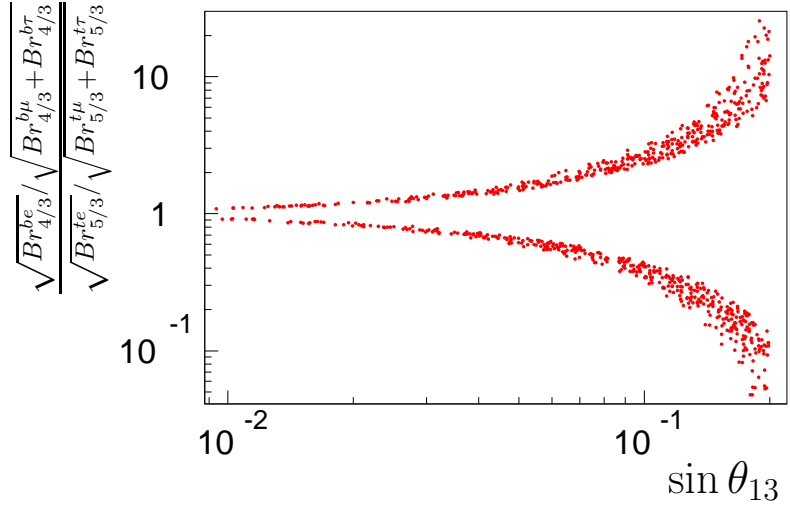}
  \caption[Ratio of decay branching ratios versus
  $\sin\theta_{13}$]
  {Ratio of decay branching ratios
    $\sqrt{Br_{4/3}^{be}}/\sqrt{Br_{4/3}^{b\mu}+Br_{4/3}^{b\tau}}/
    \sqrt{Br_{5/3}^{te}}/\sqrt{Br_{5/3}^{t\mu}+Br_{5/3}^{t\tau}}$
    versus $\sin\theta_{13}$.}
  \label{fig:b43cob53c_s13}
\end{figure}

Numerically we have found that there is certain combination of ratios
of branching ratios that is correlated with $\sin\theta_{R}=\sin\theta_{13}$
as shown in figure \ref{fig:b43cob53c_s13}. With the current upper limit on
$\sin\theta_{R}$, this ratio is not very much constrained. However,
a future measurement of $\sin\theta_{R}$, would confine this ratio
to lie in a very small, albeit double-valued, interval and thus such a
measurement could become a powerful experimental cross-check of the
scenario discussed here. Note also that this ratio approaches $1$ for
small values of $\sin\theta_{R}$, thus also an improved upper limit on
this angle will lead to an interesting constraint.

\begin{figure}[h!]
  \centering
  \includegraphics[width=10cm,height=7cm]{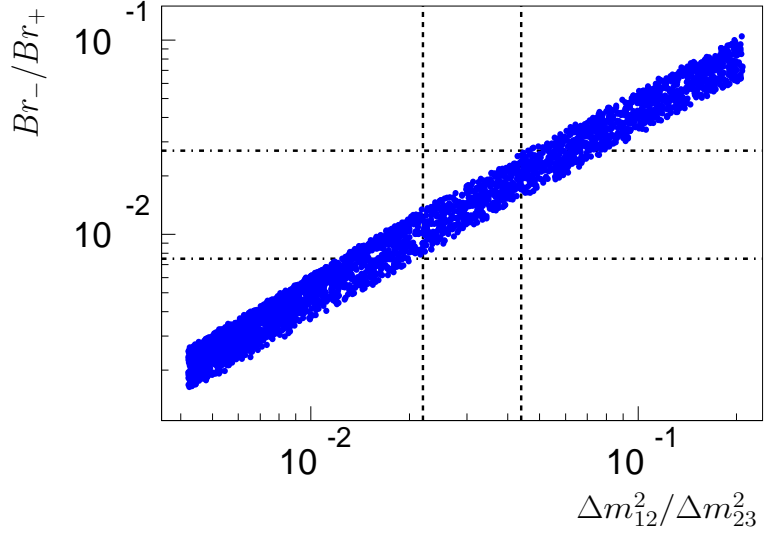}
  \caption[$B_-/B_+$ versus $R$]
   {Ratio of decay branching ratio $Br_-/Br_+$ versus $R$. Vertical
    lines indicate current $3\sigma$ range for $R$ whereas horizontal
    lines show the predicted range for this observable.
    The spread of the points in the plot determine the uncertainity with
    which this ratio can currently be predicted. }
  \label{fig:brcom_mn2sqmn3sq}
\end{figure}

Finally, from equation (\ref{eq:ratio}) one expects that
\begin{eqnarray}
  \label{eq:rat_mn1mn2}
  R=\frac{Br_- '}{Br_+ '}&\equiv&
  \frac{\sqrt{\sum_{i=e,\mu\tau}Br_{5/3}^{t i}Br_{4/3}^{b i}}
    - \sqrt{\sum_{i,j=e,\mu\tau}Br_{5/3}^{t i}Br_{4/3}^{b j}}}
  {\sqrt{\sum_{i=e,\mu\tau}Br_{5/3}^{t i}Br_{4/3}^{b i}}
    + \sqrt{\sum_{i,j=e,\mu\tau}Br_{5/3}^{t i}Br_{4/3}^{b j}}} \\ \nonumber
&\simeq &
  \frac{\sqrt{\sum_{i=\mu\tau}Br_{5/3}^{t i}Br_{4/3}^{b i}}
    - \sqrt{\sum_{i,j=\mu\tau}Br_{5/3}^{t i}Br_{4/3}^{b j}}}
  {\sqrt{\sum_{i=\mu\tau}Br_{5/3}^{t i}Br_{4/3}^{b i}}
    + \sqrt{\sum_{i,j=\mu\tau}Br_{5/3}^{t i}Br_{4/3}^{b j}}}.
\end{eqnarray}
The neglection of electron final states in the 2nd equation above
is motivated by eq. (\ref{eq:b4353comrat}). Numerical results are
shown in figure \ref{fig:brcom_mn2sqmn3sq}. The spread of the points
in the plot gives the precision with which the ratio $Br_-/Br_+$ can
be predicted, neglecting electron final states and scanning over the
allowed ranges of other neutrino physics observables. As demonstrated
by figure \ref{fig:brcom_mn2sqmn3sq} the observable $Br_-/Br_+$ is
currently expected to lie within the range
$[7.5\times 10^{-3},2.9\times 10^{-2}]$.

All results shown in the figures (\ref{fig:br_t23sq})-
(\ref{fig:brcom_mn2sqmn3sq}) are based on the assumption that the
top-loop gives the most important contribution to the neutrino mass
matrix. However, very similar results can be obtained if the bottom
loop dominates in either scenario II.a or scenario II.b. We will
not repeat the discussion in detail here. The results for scenario II.a
can be obtained by the replacement of
${\widehat S}_{5/3} \rightarrow {\widehat S}_{2/3}$
and scenario II.b by the replacements
${\widehat S}_{5/3} \rightarrow {\widehat S}_{2/3}$
and
${\widehat S}_{4/3} \rightarrow {\widehat S}_{1/3}$
in all equations and figures above.

In summary, qualitative expectations for some ratios of branching ratios
of fermionic LQ decays can be derived from the requirement that LQ loops
explain neutrino oscillation data. In general, lepton flavour violating
decays with similar branching ratios to muonic and tau final states are
expected for some specific LQ decays. Sharp predictions for various decay
modes can be made, under the reasonable asumption that one LQ loop dominates
over all others.

\subsection{Leptoquark decays to Higgs and gauge boson final states}

Since the current lower limit on the mass of a standard model
like Higgs boson is $m_{h^0}\ge 114.4$ GeV \cite{Yao:2006px},
one expects that LQs can decay also to standard model gauge bosons,
$W^{\pm}$ and $Z^0$, if the Higgs final state is kinematically possible.
We will therefore discuss partial decay widths to Higgs, $W^{\pm}$
and $Z^0$ final states jointly in this subsection.

In the model discussed here, heavier LQs can decay to lighter LQs
plus a standard model Higgs boson, i.e. $(\widehat{S}_{Q})_j\to h^0 +
(\widehat{S}_{Q})_i$, due to the interactions given in eq.
(\ref{eq:LQhiggsint}). Partial decay widths can be written as
\begin{equation}
  \label{eq:dwhiggs}
  \Gamma[(\widehat{S}_{Q})_j\to h^0 + (\widehat{S}_{Q})_i]
  =\frac{1}{16\pi}\widetilde{g}_Q^2\,m_{S_j}\lambda^{1/2}
  (1,r_{ij},r_h) .
\end{equation}
Here, the arguments of $\lambda^{1/2}(a,b,c)$ have been defined dimensionless,
$r_{ij}\equiv m_{S_{i}}^2/m_{S_j}^2$ and $r_{h}\equiv m_{h^0}^2/m_{S_j}^2$.
The effective couplings $\widetilde{g}_Q$ for the different values of
$Q=-1/3,-2/3,-4/3,-5/3$ are defined as
\begin{eqnarray}
  % q=-1/3 effective coupling
  \widetilde{g}_{-1/3} = & &
  \frac{g^{(LR)}_{S_0}}{2}\frac{v}{m_{S_j}}
  R^{1/3}_{j1}R^{1/3}_{i2} +
  \frac{h^{(L)}_{S_0}}{\sqrt{2}\,m_{S_j}}
  R^{1/3}_{j1}R^{1/3}_{i3} +
  \frac{\kappa^{(L)}_S}{2}\frac{v}{m_{S_j}}
  R^{1/3}_{j1}R^{1/3}_{i4}
  \nonumber\\
  & + &\frac{h^{(R)}_{S_0}}{\sqrt{2}\,m_{S_j}}
  R^{1/3}_{j2}R^{1/3}_{i3}
  +\frac{\kappa^{(R)}_S}{2}\frac{v}{m_{S_j}}
  R^{1/3}_{j2}R^{1/3}_{i4} +
  \frac{h_{S_1}}{\sqrt{2}\,m_{S_j}}
  R^{1/3}_{j3}R^{1/3}_{i4}
\end{eqnarray}
\begin{eqnarray}
  % q=-2/3 effective coupling
  \widetilde{g}_{-2/3} = & &
  \frac{Y_{S_{1/2}}^L}{2}\frac{v}{m_{S_j}}
  R^{2/3}_{j1}R^{2/3}_{i2} +
  \frac{Y_{S_{1/2}}^R}{2}\frac{v}{m_{S_j}}
  R^{2/3}_{j1}R^{2/3}_{i3} +
  \frac{h_{S_1}}{m_{S_j}}R^{2/3}_{j1}R^{2/3}_{i4}
  \nonumber\\
  & +& \frac{g^{(LR)}_{S_{1/2}}}{2}\frac{v}{m_{S_j}}
  R^{2/3}_{j2}R^{2/3}_{i3} ,
\end{eqnarray}
\begin{equation}
  % q=-4/3 effective coupling
  \widetilde{g}_{-4/3} =
  \frac{Y_{S_1}}{\sqrt{2}}\frac{v}{m_{S_j}}
  R^{4/3}_{j1}R^{4/3}_{i2} ,\\
\end{equation}
\begin{equation}
  % q=-5/3 effective coupling
  \widetilde{g}_{-5/3} =
  \frac{g^{(LR)}_{S_{1/2}}}{2}\frac{v}{m_{S_j}}
  R^{5/3}_{j1}R^{5/3}_{i2} .
\end{equation}
$R^Q_{ij}$ are the rotation matrices, which diagonalize the LQ mass
matrices. Note, that the above couplings contain the same parameters
which induce neutrino masses due to LQ mixing.

For any given set of LQs of charge $Q$ the couplings with the
$Z^0$ can be written as
\begin{equation}
  \label{eq:kinterms3}
    \frac{ig}{\cos\theta_W}
    Z^\mu
    \sum_l\left(T_3^l-Q\;\sin^2\theta_W\right)
      S_Q^l \overleftrightarrow{\partial}_\mu (S_Q^l)^\dagger .
\end{equation}
Non-diagonal couplings of the $Z^0$ gauge boson to different LQ states
of the same $Q$, but different $T_3$ appear, after rotation to the
mass eigenstate basis. The partial decay width can be written as
\begin{equation}
  \label{eq:decaywidth}
    \Gamma[(\widehat{S}_{Q})_j\to Z^0 + (\widehat{S}_{Q})_i]
   =\frac{1}{16\pi}\frac{g^2}{\cos\theta_W^2}
  \theta_Q^2\;\frac{M_{S_j}^3}{M_Z^2}
  \lambda^{3/2}(1,r_{ij},r_Z) ,
\end{equation}
where $r_Z = (m_{Z^0}/m_j)^2$ and
\begin{eqnarray}
  \label{eq:mixfactorsZcoupl}
  \theta_{-1/3}&=&-\frac{1}{2}R^{1/3}_{j3}
  R^{1/3}_{i3}\\ \nonumber
  \theta_{-2/3}&=&-(R^{2/3}_{j1}R^{2/3}_{i1}
  + \frac{3}{2}R^{2/3}_{j4}R^{2/3}_{i4})\\ \nonumber
  \theta_{-4/3}&=&-R^{4/3}_{j2}R^{4/3}_{i2}.
\end{eqnarray}
Note that $Q=-5/3$ LQs do not have any decays to $Z^0$ bosons, since
their couplings to $Z^0$ are completely diagonal. Closer inspection
of eq. (\ref{eq:mixfactorsZcoupl}) reveals that the decays to $Z^0$
states can occur only if LQ mixing (by the same parameters which
govern the Higgs final states) is non-zero. Thus, also observation
of $Z^0$ final states gives valuable information about the parameters
in eq. (\ref{eq:LQhiggsint}).

Heavier LQs can decay to a lighter one and a $W^{\pm}$ gauge boson,
$S_Q\to W + S_{Q'}$, where $S_Q$ and $S_{Q'}$ are members
of the same doublet (triplet). Possible decays therefore are:
\begin{align}
  \label{eq:gboson1}
  (\widehat{S}_{-5/3})_j&\leftrightarrow W^{-} + (\widehat{S}_{-2/3})_i ,\\
  \label{eq:gboson2}
  (\widehat{S}_{-2/3})_j&\leftrightarrow W^{-} +(\widehat{S}_{-1/3})_i^\dagger ,\\
  \label{eq:gboson3}
  (\widehat{S}_{-4/3})_j&\leftrightarrow W^{-} + (\widehat{S}_{-1/3})_i ,
\end{align}
where the processes in (\ref{eq:gboson1}), (\ref{eq:gboson2})
and (\ref{eq:gboson3}) come from the decays of the members of
the doublet $S_{1/2}$, $\widetilde{S}_{1/2}$ and the triplet $S_1$,
respectively, after rotation to the mass eigenstate basis.
Note that the process in eq. (\ref{eq:gboson2}) can also come
from the decay of the $T_3=1$ to the $T_3=0$ components of the
triplet. The decay widths for the processes in (\ref{eq:gboson1}),
(\ref{eq:gboson2}) and (\ref{eq:gboson3}) can be written as
\begin{equation}
  \label{eq:gbw1}
  \Gamma[(\widehat{S}_{Q})_j\to W^{\pm} + (\widehat{S}_{Q'})_i]
  =\frac{g^2\theta_Q^2}{32\pi} \frac{m_{S_j}^3}{M_W^2}
  \lambda^{3/2}(1,r_{ij},r_W) .
\end{equation}
Here $r_{W}\equiv M_W^2/m_{S_j}^2$ and the mixing factors
are given by
\begin{align}
  \label{eq:mix1}
  \theta_{-5/3}&=(R^{2/3})_{i2}(R^{5/3})_{j1}+
  (R^{2/3})_{i3}(R^{5/3})_{j2} ,\\
  \label{eq:mix2}
  \theta_{-2/3}&=(R^{1/3})_{i3}(R^{2/3})_{j1}+
  \sqrt{2}(R^{1/3})_{i4}(R^{2/3})_{j4}\\
  \label{eq:mix3}
  \theta_{-4/3}&=\sqrt{2}(R^{1/3})_{i4}(R^{4/3})_{j2} .
\end{align}
Our formula eq. (\ref{eq:gbw1}) agrees with the one calculated earlier
in \cite{Doncheski:1997bq}, once LQ mixing is properly taken
into account.

We now turn to a discussion of typical ranges for the branching
ratios of bosonic final states. We will first discuss the
example of decays of LQs with $Q=4/3$, assuming the decay
to $Q=1/3$ LQs plus $W^{\pm}$ is kinematically closed.

\begin{figure}[h!]
\begin{center}
\includegraphics[width=60mm,height=40mm]{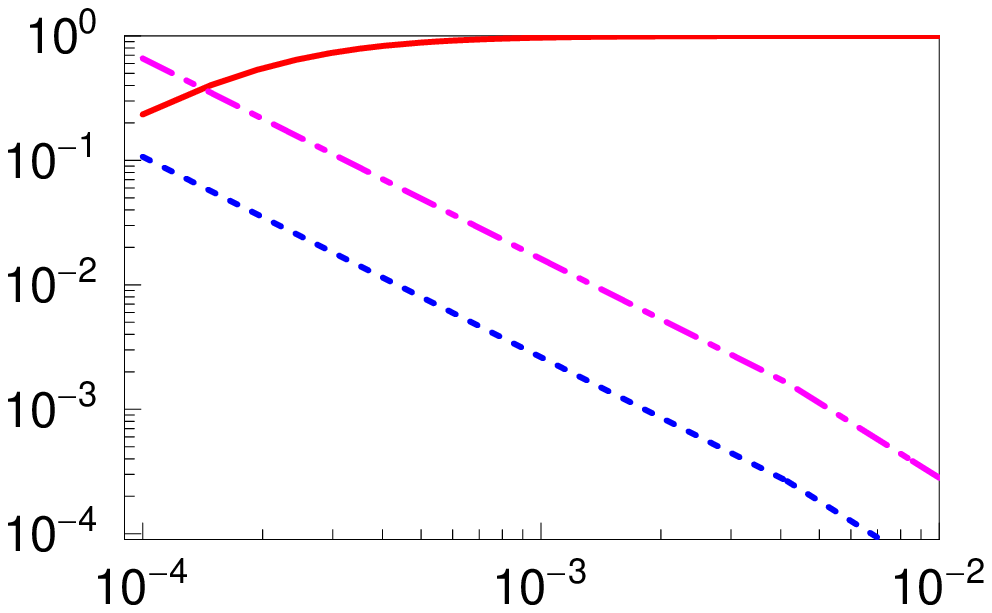}
\hskip10mm
\includegraphics[width=60mm,height=40mm]{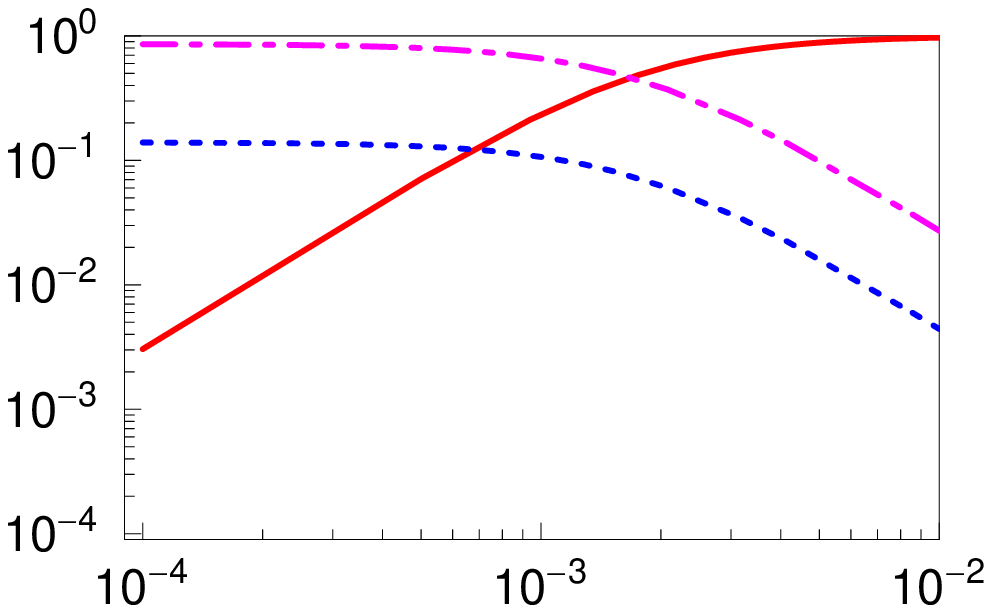}
\vskip4mm
\includegraphics[width=60mm,height=40mm]{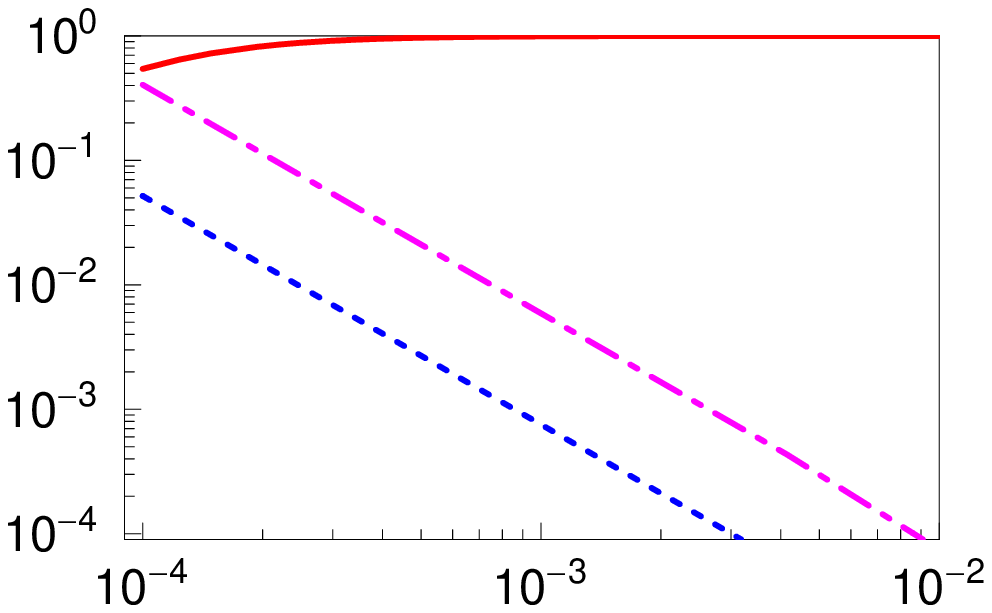}
\hskip10mm
\includegraphics[width=60mm,height=40mm]{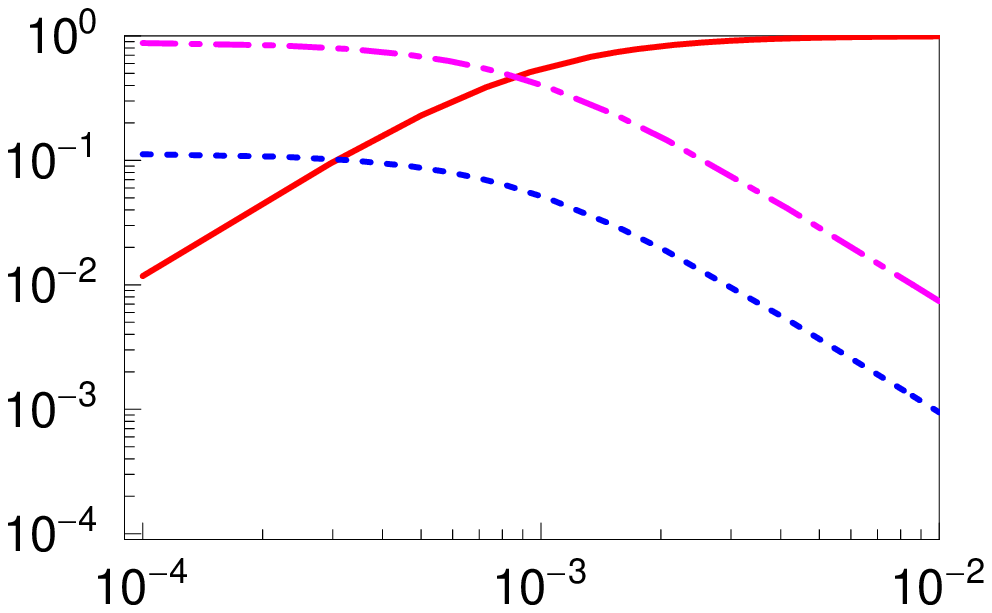}
\end{center}
\caption{Typical values for decay branching ratios for the heavier
$Q=4/3$ LQ, in case $W^{\pm}$ final states are kinematically closed.
Branching ratios are plotted versus the ``average''
Yukawa coupling ${\bar\lambda}\equiv\sqrt{\sum_{i}\lambda_{i3}^2}$,
for different values of $Y_{S_1}$ and $m^{Q=4/3}_{2}$. Full line
(red): fermionic final states, dashed line (blue) Higgs final
state, dot-dashed (magenta) $Z^0$ final state. In all figures
$m^{Q=4/3}_{1}$ has been set to $m^{Q=4/3}_{1}= 250$ GeV and
we have chosen $m_{h^0} = 115$ GeV, motivated by the LEP limit. Top left:
($Y_{S_1}=0.01, m^{Q=4/3}_{2}=400$ GeV); top right: ($Y_{S_1}=0.1,
m^{Q=4/3}_{2}=400$ GeV); bottom left:
($Y_{S_1}=0.01, m^{Q=4/3}_{2}=800$ GeV); bottom right: ($Y_{S_1}=0.1,
m^{Q=4/3}_{2}=800$ GeV).}
\label{fig:lq43br}
\end{figure}

Fig. (\ref{fig:lq43br}) shows a set of numerical examples of
branching ratios of the heavier of the $Q=4/3$ LQ mass eigenstate
to fermionic, $h^0$ and $Z^0$ final states, for some typical
choices of parameters, see figure caption. For Yukawa
couplings of the order ${\bar\lambda}\sim 10^{-3}$ values of
$Y_{S_1}$ as small as $Y_{S_1}\simeq 10^{-2}$ can lead to
observable branching ratios into bosonic final states.

While we can not predict whether fermionic or bosonic final states
will dominate, it is interesting to note that the current LQ model
makes a definite prediction for the {\em ratio of branching ratios}
of $h^0$ and $Z^0$ final states, if $m^{Q=4/3}_{2}$ is sufficiently
larger than $m^{Q=4/3}_{1}+m_{h^0}$. This can be understood as
follows. If $m^{Q=4/3}_{2}$ is much larger than $m^{Q=4/3}_{1}$,
$m_{h^0}$ and $m_{Z^0}$, one can neglect the phase space correction
factors, $\lambda(1,x,y)$, and approximate the $Q=4/3$ mixing angle by
$\theta_{Q=4/3} \simeq \sqrt{2}Y_{S_1}v^2/(m^{Q=4/3}_{2})^2$.
The ratio of the partial widths to Higgs and $Z^0$ states is
then simply given by
\begin{equation}
\label{eq:ratiopredZ}
\frac{\Gamma[(\widehat{S}_{4/3})_2\to Z^0 + (\widehat{S}_{4/3})_1]}
{\Gamma[(\widehat{S}_{4/3})_2\to h^0 + (\widehat{S}_{4/3})_1]}
\simeq
4 \frac{g^2}{c_W^2}\frac{v^2}{m_{Z^0}^2} \sim 8,
\end{equation}
independent of all non-SM parameters. This explains the ratio
observed in the numerical examples of fig. (\ref{fig:lq43br})
and constitutes a nice consistency test for the LQ model of
neutrino masses.

We now turn to $W^{\pm}$ final states. In general, in electro-weak
symmetry breaking new mass terms for LQs, which are members of the
same multiplet, could be generated by some non-SM scalars, potentially
introducing large splitting {\em within} a given multiplet. In this
case, LQ decays to $W^{\pm}$ states could occur independent of LQ
mixing between {\em different} multiplets. Then, since LQ-$W^{\pm}$
decays are of order $g^2$ they would easily become dominant once
kinematically allowed. In the current model, however, mass splitting
of LQs within the same multiplet comes only from LQ mixing, see
eqs (\ref{eq:q1/3lqmassm})-(\ref{eq:q5/3lqmassm}). Thus LQ-$W^{\pm}$
final states should have widths similar to the $Z^0$ final states
discussed above. Consider, for example the decays of a $Q=5/3$ LQ.
The mass matrix of the $Q=5/3$ LQs, see eq. (\ref{eq:q4/3lqmassm}),
contains the same parameters as a (2-by-2) submatrix of the
$Q=2/3$ mass matrix, compare to eq. (\ref{eq:q1/3lqmassm}).
If the other $Q=2/3$ states are heavier than these states, we can
give a similar estimate of the ratio of Higgs and $W^{\pm}$ final
states, as has been derived above for $Z^0$ final states, see
eq. (\ref{eq:ratiopredZ}). Assuming again $m^{Q=5/3}_2$ being
much heavier than final state particles, we find
\begin{equation}
\label{eq:ratiopredW}
\frac{\Gamma[(\widehat{S}_{5/3})_2\to h^0 + (\widehat{S}_{5/3})_1]}
{\Gamma[(\widehat{S}_{5/3})_2\to W^{\pm} + (\widehat{S}_{2/3})_1]}
\simeq
\frac{1}{8 g^2}\frac{m_W^2}{v^2} \simeq 0.063.
\end{equation}
In the general situation, however, when all $Q=2/3$ states are
relatively light, the branching ratio to $h^0$ final states can
not be predicted accurately. Thus, $W^{\pm}$ final states can not
provide an as valuable test for the model as is the case for $Z^0$
decays.

In summary, heavier LQs will decay to bosonic final states, if
kinematically allowed. Since in the current model all these
decays are induced by the presence of the LQ-Higgs interaction
parameters, observing such decays are an essential test of the
LQ model of neutrino mass. Branching ratios for bosonic final
states typically fall into the range ${\cal O}(10^{-4}-1)$,
for LQ-Higgs couplings order ${\cal O}(10^{-2}-1)$. Although
we have discussed only the cases $Q=4/3$ and $Q=5/3$, bosonic
widths of LQs with other electric charges are expected to show
a very similar parameter dependence (and therefore similar branching
ratios).

\section{Summary}
\label{sec:LQcon}
LQ fields with baryon number conserving Yukawa interactions can have
masses at or near the electro-weak scale. If these LQ fields couple
to the SM Higgs, the resulting model generates neutrino masses at the
one-loop level. In this work we have explored the phenomenological
consequences of LQs as the origin of the observed neutrino masses for
future accelerator experiments, such as the LHC.

Fermionic decays of (some of the) LQ states trace the neutrino angles,
i.e. certain ratios of decay branching ratios can be predicted from
current neutrino data. In general one expects that those LQs, which give
the dominant contribution to the neutrino mass matrix, if (pair) produced
at the LHC decay with sizeable flavour violation. For these states there
should be a similar number of events with $\tau^{\pm}\mu^{\mp}$ final
states, as there are final states with muon and tau pairs. One also
expects a smaller number of events of the type $e^{\pm}\mu^{\mp}$
(and $e^{\pm}\tau^{\mp}$), although the details in this case are more
involved, as discussed above.

In this context we would like to stress that one of the basic assumptions
applied in practically all accelerator searches for LQs is that LQs couple
only to one generation of leptons and quarks at a time. As discussed at
length above, such completely generation diagonal couplings would
{\em exclude} LQ-loops as an explanation of neutrino oscillation data.
Extending the LQ search to lepton flavour violating decays thus should
be considered seriously by experimentalists.

We have also discussed how, in some specific scenarios, much more
detailed predictions can be made. Given the observed hierarchy of
standard model quark masses, it seems reasonable to assume that
contributuins from 3rd generation quark loops dominate the neutrino
mass matrix. For the case of top quark dominance, our results are
summarized in the figures (\ref{fig:br_t23sq})-
(\ref{fig:brcom_mn2sqmn3sq}). Similar results hold in case of
pure bottom-loop dominance.

Finally, an important test of the hypothesis that LQs can generate Majorana
neutrino masses, is the search for decays of heavier LQs into lighter
ones plus a standard model Higgs or gauge boson. Any observation of a
non-zero branching ratio for the decay $S_i \to S_j + h^0/Z^0$ constitutes
proof for LQ mixing, which is the basic ingredient for the LQ explanation
of neutrino masses. If LQs are found at the LHC, the search for such
decays should be made a priority.

\section*{Acknowledgments}
D.A.S wants to thanks the ``{\it Instituto de F\'{i}sica de la Universidad
de Antioquia}'' for their hospitality.
This work was supported by Spanish grant FPA2005-01269, by the European
Commission Human Potential Program RTN network MRTN-CT-2004-503369 and 
by Chilean grant~ CONICYT PBCT/No.285/2006. D.A.S. is supported by a 
Spanish PhD fellowship by M.C.Y.T.

\end{document}